\documentstyle[11pt,aaspp4]{article}

\def\kms{km s$^{-1}$}

\def\mh{M_{\bullet}}
\def\rhl{\rho_{\bullet,L}}
\def\rhz{\rho_{\bullet,z}}
\def\mb{M_{\rm bulge}}
\def\ms{\mh-\sigma}
\def\msun{M$_{\odot}$}
\def\mpc{M$_{\odot}$ Mpc$^{-3}$}

\lefthead{Merritt \& Ferrarese}
\righthead{Black Hole Demographics}

\begin{document}

\title{Black Hole Demographics from the $\mh-\sigma$ Relation}

\author{David Merritt and Laura Ferrarese}  
\affil{Rutgers University, New Brunswick, NJ, 08854}
\authoraddr{Department of Physics and Astronomy, 136 Frelinghuysen Road, 
Piscataway, NJ 08854}

\vspace{.3in}


\vspace{.2in}

\begin{abstract}

We analyze a sample of 32 galaxies for which a dynamical estimate of
the mass of the hot stellar component, $\mb$, is available. For each
of  these galaxies, we calculate the mass of the central black hole,
$\mh$, using the tight empirical correlation between $\mh$ and the bulge 
stellar velocity dispersion. The frequency function  $N[\log (\mh/\mb)]$ is
reasonably well described as a Gaussian with
$\langle\log(\mh/\mb)\rangle \approx -2.90$  and standard deviation
$\sim 0.45$; the implied mean ratio of black hole to bulge mass is a
factor $\sim 5$ smaller than generally quoted in the literature. We
comment on marginal evidence for a lower, average black-hole mass
fraction in more massive galaxies, which should be investigated using
larger samples. The total mass density in BHs in the local Universe is
estimated to be $\sim 5 \times 10^5$ \mpc, consistent with that
inferred from high redshift ($z \sim 2$) AGNs.

\end{abstract}

\section{Introduction}

With an ever-increasing number of secure detections, supermassive
black holes (BHs) have evolved, in a ten-year span, from exotic
curiosities to fundamental components of galaxies.  It is  now
generally accepted that the formation and evolution of  galaxies and
supermassive BHs are tightly intertwined, from  the early phases of
proto-galactic formation (Silk \& Rees 1998;  Sellwood \& Moore 1999),
through hierarchical build-up in CDM-like  cosmogonies (Efstathiou \&
Rees 1988; Haehnelt \& Rees 1993; Haehnelt, Natarajan \& Rees 1998;
Haiman \& Loeb 1998),  to recent galaxy mergers (Merritt 2000).
Studying the demographics of the local  BH population might have a
significant impact on models of galaxy evolution  (e.g. Salucci et
al. 1999; Cattaneo, Haehnelt \& Rees 1999;  Kauffmann \& Haehnelt
2000).

Magorrian et al.  (1998) presented the first, and to date only,
demographic study of nuclear BHs.  Ground based kinematic data for 32
galaxies were combined with HST photometry to constrain dynamical
models -- based on the Jeans equation -- under the assumptions of
axisymmetry, velocity isotropy in the meridional plane and a
spatially-constant mass-to-light ratio for the stars.  The mass of a
putative nuclear BH was introduced as a free parameter, in addition to
the stellar mass-to-light ratio and the galaxy inclination  angle.  In
most of the galaxies, the addition of a central point mass improved
the fit to the observed kinematics.  Magorrian et al.  concluded that
most, or all, galaxies contain central BHs with an average  ratio of
BH mass to spheroid mass of $\mh/\mb \sim 10^{-2}$.

The Magorrian et al.  study remains unique for targeting a large
sample of galaxies, and for its  coherent and homogeneous treatment of
the data.  However, while  the Magorrian et al.  estimates of the
bulge mass-to-light ratios are likely to be robust, a number of
authors have noted that the inferred BH masses might be systematically
too  large.  Van der Marel (1997) showed that the BH masses  derived
from well-resolved central kinematical data are a factor 5 smaller
than produced by the Magorrian et al. analysis; he suggested that the
neglect  of velocity anisotropy might have led to overestimates of the
BH  masses. Wandel (1999)  compared BH masses derived from
reverberation mapping studies of  active galaxies with the Magorrian
et al.  estimates and found a  discrepancy of a factor of $\sim 20$ in
the BH-to-bulge mass ratio at  a fixed luminosity.  He noted the
difficulty of resolving low-mass BHs  in distant galaxies and
suggested a distance-dependent bias in the  estimates.

An independent argument along the same lines was presented by
Ferrarese  \& Merritt (2000; FM00; Paper I).  Using the tight empirical
correlation between  $\mh$ and $\sigma$, the velocity dispersion of
the stellar bulge, for  the 12 galaxies with the best-determined BH
masses, FM00 showed that  the Magorrian et al.  masses fall
systematically above the $\ms$  relation, some by as much as two
orders of magnitude.

At the present time, the $\ms$ relation is probably our best  guide to
BH demographics.   Ferrarese \& Merritt (2000) found that the relation
has a scatter  no larger than that expected on the basis of
measurement errors  alone.  The relation is apparently so tight that
it surpasses in  predictive accuracy what can be achieved from
detailed dynamical  modeling of stellar kinematical data in most
galaxies.  By  combining the bulge stellar masses derived by Magorrian
et al.   with BH masses inferred from the $\ms$ relation, we are in a
position  to compute the most robust estimate to date of the BH mass
distribution  in nearby galaxies.
 
\section{Data}

Table 1 gives the relevant physical parameters for the 32 galaxies  in
the Magorrian et al. sample. All galaxies, with the exception of M31,
are early type. In what follows, we refer to the hot stellar component
in these galaxies as the ``bulge;'' this is in fact the case for M31,
although for the other objects the ``bulge'' is the entire galaxy.
Distances were re-derived as in Paper I; values for the bulge $V$-band
luminosity ($L_{\rm bulge}$),  bulge mass ($M_{\rm bulge}$) and BH
mass ($M_{\rm fit}$)  are the same as in Magorrian et al. except for
the (mostly small) corrections resulting from the new distances.

Central velocity dispersions $\sigma_c$ were taken from the literature
and corrected to a common aperture size of $1/8$ of the effective
radius,  as in Paper I.  We then computed BH masses, $\mh$, using the
$\mh-\sigma_c$ relation  in the form given by Merritt \& Ferrarese
(2000; Paper II):
\begin{equation}
\mh = 1.30 \times 10^8~{\rm M}_{\odot}(\sigma_c/200\ {\rm km\
s}^{-1})^{4.72}.
\end{equation}
This expression was derived by fitting to the combined galaxy samples
of Ferrarese \& Merritt (2000) (12 galaxies) and Gebhardt et
al. (2000a) (15 additional galaxies), plus 7 active galaxies for which
both $\sigma_c$ and $\mh$  are available, the latter from
reverberation mapping  (Gebhardt et al. 2000b).  Some debate exists
over the exact value of the slope in equation (1)  (Paper I, II;
Gebhardt et al. 2000a). We explore  below how changing the assumed
slope affects our conclusions.

The correlations between $\mh$ and $L_{\rm bulge}$, and between $\mh$
and $M_{\rm bulge}$, are shown in Figure 1.  There is a rough
proportionality of both $L_{\rm bulge}$ and $M_{\rm bulge}$ with
$\mh$, though the vertical scatter in both relations is much larger
than in the $\ms$ relation (Paper I).

We defined the two mass ratios:
\begin{eqnarray}
x_{\rm fit} & \equiv & M_{\rm fit}/M_{\rm bulge}, \nonumber \\ x &
\equiv & \mh/M_{\rm bulge}
\end{eqnarray}
based respectively on the BH mass estimates from Magorrian et al.  and
from the $\ms$ relation.  Values of $\log x_{\rm fit}$ and $\log x$
are given in Table 1.  BH masses derived from the $\ms$ relation yield
the mean values $\langle x\rangle=2.50\times 10^{-3}$ and $\langle\log
x\rangle =-2.90$.  These are substantially smaller than the mean
values computed from  the Magorrian et al. BH masses: $\langle x_{\rm
fit}\rangle=1.68\times 10^{-2}$ and  $\langle \log x_{\rm
fit}\rangle=-2.20$.  We note that one galaxy, NGC 4486b, has $\log
x_{\rm fit}=-0.54$, making it an extreme outlier in the Magorrian et
al. distribution.  Removing this single galaxy from the sample gives
$\langle x_{\rm fit}\rangle=7.2\times 10^{-3}$ while leaving $\langle
\log x_{\rm fit}\rangle$ essentially unchanged.

Figure 2 reveals a clear trend of $M_{\rm fit}/\mh$ with the  apparent
radius of influence of the central black hole, assuming the masses
predicted by the $\ms$ relation are correct.  A natural interpretation
is that there is a resolution-dependent  bias in the Magorrian et
al. modeling (e.g. van der Marel 1997; Wandel 1999): the radius of
influence of all of the  Magorrian et al. galaxies is smaller than $1$
arcsec, too small to have been clearly resolved from the ground.

\section{Analysis}

We seek an estimate of the frequency function $N(y)=N(\log x)$.
Following Merritt (1997), we define this estimate as $\hat N(y)$,  the
function that maximizes the penalized log likelihood
\begin{equation}
\log{\cal L}_p = \sum_{i=1}^n \log (N\circ E)_i - \lambda P(N)
\end{equation}
of the data $y_i, i=1,...,n$, subject to the constraints
\begin{equation}
\int N(y) dy = 1, \ \ \ \ N(y) \ge 0.
\end{equation}
Here $N\circ E$ is the ``observable'' function, i.e.  the convolution
of the true $N$ with the error distribution of $y$.  This error
distribution is not well known;  we assume that it is a Gaussian with
some dispersion $\Delta_y$.  Failing to account for measurement errors
in $y$  would lead to a spuriously broad $\hat N(y)$.

The natural penalty function to use is Silverman's (1982):
\begin{equation}
P(N) = \int_{-\infty}^{+\infty}\left[(d/dy)^3\log N(y)\right]^2 dy.
\end{equation}
This function assigns zero penalty to any $N(y)$ that is Gaussian.  In
the limit of large $\lambda$,  the estimate $\hat N$ is driven toward
the Gaussian function that is  most consistent, in a
maximum-likelihood sense, with the data; smaller values of $\lambda$
return nonparametric estimates of $N(y)$.  We made no attempt to
calculate the ``optimal'' value of the smoothing parameter given the
small size of the data set.

The results of the optimization are shown in Figure 3, assuming
$\Delta_{y}= 0.15$.  $\hat N(y)$ is nicely symmetric and reasonably
well described as a Gaussian, although with a narrower-than-Gaussian
central peak.  The best-fit Gaussian has its mean at  $y = \log
x=-2.93$ and a standard deviation of $0.45$.

By contrast, the Magorrian et al. masses define a more flat-topped
distribution with one extreme outlier,  NGC 4486b, at $\log x_{\rm
fit}=-0.54$.  The Gaussian fit to the Magorrian et al. mass
distribution has its mode at $ -2.25$ and a standard deviation of
$0.52$.

The two galaxies with the largest BH mass ratios, NGC 4486b and NGC
4660, are both low-mass ellipticals.  The smallest mass ratio, $\log
x=-3.95$, is seen in a very massive galaxy, NGC 4874.  It is therefore
interesting to check whether low- and high-mass galaxies have
different characteristic distributions of $\log x$.  This hypothesis
is tested in Figure 4a,  which shows $\hat N(y)$ computed separately
for the 16 galaxies from Table 1 with the lowest and highest values of
$\mb$.  There is in fact a slight difference between the two
distributions: the high-mass galaxies have  $\langle\log x\rangle =
-3.10$ and $\sigma_{\log x}=0.39$, while the low-mass galaxies have
$\langle\log x\rangle = -2.71$ and $\sigma_{\log x}=0.49$.  However
the offset in $\langle\log x\rangle$ is similar to the width of either
distribution and may not be significant.  We note that the massive
galaxies define a narrower distribution.

Our conclusions about $N(y)$ might be substantially dependent on the
assumed form of the $\ms$ relation, equation (1).  The slope of that
relation is fairly uncertain, $\alpha=4.72\pm0.4$; however the
normalization at $\sigma_c\approx 200$  \kms appears to be more robust
(Paper II).  We therefore set
\begin{equation}
\mh = 1.30 \times 10^8~{\rm M}_{\odot}(\sigma_c/200\ {\rm km\
s}^{-1})^{\alpha}
\end{equation}
and investigated the effects of varying $\alpha$.  Figure 4b shows
that a larger slope implies a broader $N(y)$ due to the stronger
implied dependence of $\mh$ on $\sigma_c$.  However the mean value of
$\log x$ is almost unchanged.

\section{Discussion}

Our estimate of the mean BH-to-bulge mass ratio, $\langle\log
x\rangle\approx -2.90$, falls squarely between the estimates of
Magorrian et al. (1998) ($\sim -2.28$), based on dynamical modeling of
the same sample of galaxies  used here; and of Wandel (1999) ($\sim
-3.50$), based on BH masses computed from reverberation mapping in  a
sample of 18 active galaxies.

Bulge masses in the Wandel (1999) study were computed directly from
bulge luminosities assuming a simple scaling law for the mass-to-light
ratio, and not from dynamical modeling.  There is reason to believe
that these luminosities are systematically  too large and therefore
that the derived mass ratios $\mh/\mb$ are too low.  Gebhardt et
al. (2000b) and Merritt \& Ferrarese (2000) found that the
reverberation mapping BH masses in 7 galaxies were consistent with the
$\ms$ relation even though they fall systematically below the
$\mh-L_{\rm bulge}$ relation.  A reasonable conclusion is that the
true or derived luminosities of these active galaxies are
systematically higher than those of normal galaxies with comparable
velocity dispersions.  A mean offset of a factor $\sim 4$ in the bulge
luminosities  would suffice to bring the average mass ratio for active
galaxies in line with the value inferred here.  Gebhardt et
al. (2000b) discuss a number of possible reasons why an error of this
sort is likely in the AGN bulge luminosities.

The discrepancy with the Magorrian et al. (1998) masses is perhaps
unsurprising given past indications that these masses are
systematically too large (van der Marel 1997; Ho 1999). The difference
between  $\langle\log x\rangle$ and $\langle\log x_{\rm fit}\rangle$
corresponds  to a factor $\sim 5$ average error in the Magorrian et
al. BH masses.  One possible explanation is the neglect of anisotropy
in the modeling (van der Marel 1997), but we emphasize that the errors
in $M_{\rm fit}$ implied by Figure 2 are enormous, of order $10-100$,
in many of the  galaxies. If the BH masses predicted by the  $\ms$
relation are correct, the kinematical data for these galaxies would
not have contained any useful information about the mass of the BH
(Figure 2).  Any features in these data that were reproduced by
adjusting $M_{\rm fit}$  must have had their origin in some systematic
difference between the models  and real galaxies, and not in the
gravitational attraction of the BH.  This conclusion, if correct,
underscores the dangers of an ``assembly-line''  approach to galaxy
modeling.

We may crudely estimate the total mass density of BHs in the local
universe by combining our result, $\mh/\mb \sim 1.3 \times 10^{-3}$,
with the mean mass density of spheroids, $\rho_{\rm bulge} \sim 3.7
\times 10^8$ \msun Mpc$^{-3}$  (Fukugita, Hogan \& Peebles 1997, for
$H_0=75$ \kms Mpc$^{-1}$).  This simple argument (first invoked by
Haehnelt, Natarajan \& Rees 1998)  gives $\rhl \sim 4.9 \times 10^5$
\msun Mpc$^{-3}$.  Salucci et al. (1999) presented a more
sophisticated treatment based on  convolution of the spheroid
luminosity function with $N(\log x)$.  They assumed a Gaussian
distribution with $\langle\log x\rangle = -2.60$ and found $\rhl \sim
1.7 \times 10^6$ \msun Mpc$^{-3}$.  Correcting their value of
$\langle\log x\rangle$ to our value of  $-2.90$ implies a factor $\sim
2$ decrease in $\rhl$,  consistent with the result of our simpler
calculation.

The total mass density of BH at large redshifts can be estimated using
an argument first suggested by Soltan (1982). Requiring the optical
QSO luminosity function to be reproduced purely by accretion onto
nuclear BHs, and assuming an accretion efficiency of 10\%,  leads to
$\rhz \sim 2 \times 10^5$ \msun Mpc$^{-3}$ (Chokshi \& Turner 1992;
Salucci et al. 1999). While independent of the cosmological model,
this result is subject to uncertainties in the  bolometric corrections
applied to the QSO magnitudes (e.g. Salucci et al. 1999), furthermore,
concerns have been raised about the completeness of the QSO luminosity
function (e.g. Goldschmidt \& Miller 1998; Graham, Clowes \& Campusano
1999). A similar argument, based on the hard X-ray background, gives
$\rhz \sim 3-4 \times 10^5$ \msun Mpc$^{-3}$ at $z \sim 1.5$  (Fabian
\& Iwasawa 1999; Salucci et al. 1999; Barger et al. 2000). These
numbers are consistent with our estimate of $\rhl$.

By contrast, $\rhz$ differs from the local BH mass density implied by
the  Magorrian et al. relation by over an order of magnitude, assuming
a canonical 10\%  accretion efficiency onto the central  black hole in
high redshift AGNs.  Haehnelt, Natarajan \& Rees (1998) and Barger et
al. (2000)  point out that if the remnants of the QSOs are to be
identified  with the BHs in present-day galaxies, the Magorrian et
al. mass distribution requires either that a large fraction of BHs
reside  within high redshift sources that are too obscured (both in
the optical  and the X-rays) to be observed, or else that a
significant amount of  accretion (with low radiative efficiency)
proceeds to the present epoch.  The need for these alternative
explanations is largely removed when the  more robust estimate of
$\rhl$ presented in this paper is adopted.

\bigskip
\acknowledgments

LF acknowledges grant NASA NAG5-8693, and DM acknowledges grants NSF
AST 96-17088 and NASA NAG5-6037.  This research has made use of the
NASA/IPAC Extragalactic Database (NED).

\clearpage

\begin{deluxetable}{lllllllll}
\tablecolumns{9}
\tablewidth{0pc}
\scriptsize
\tablenum{1}
\tablehead{
\colhead{Galaxy} &
\colhead{Distance} &
\colhead{$L_{\rm bulge}$} &
\colhead{$M_{\rm bulge}$} &
\colhead{$\sigma_c$} &
\colhead{$M_{\rm fit}$} &
\colhead{$\log x_{\rm fit}$} &
\colhead{$\mh$} &
\colhead{$\log x$} \nl
}
\startdata

N221  & 0.8$\pm$0.1  & 0.0373 & 0.00807 & 76$\pm 10$  & 0.0228& -2.53 &  0.0135 
& -2.78 \nl  
N224  & 0.8$\pm$0.0  & 0.724  & 0.350   & 112$\pm$15  & 0.598 & -2.75 &  0.0842 
& -3.62 \nl
N821  & 24.7$\pm$2.5 & 2.47   & 1.61    & 196$\pm$26  & 2.48  & -2.81 &  1.18   
& -3.13 \nl  
N1399 & 20.5$\pm$1.6 & 5.42   & 3.60    & 312$\pm$41  & 59.7  & -1.79 & 10.60   
& -2.53 \nl
N1600 & 68.5$\pm$6.6 & 19.14  & 16.90   & 307$\pm$40  & 159   & -2.05 &  9.83   
& -3.24 \nl
N2300 & 27.0$\pm$2.6 & 3.30   & 3.40    & 269$\pm$35  & 23.3  & -2.16 &  5.27   
& -2.81 \nl
N2778 & 23.3$\pm$3.4 & 0.557  & 0.359   & 171$\pm$22  &\nodata&\nodata&  0.621  
& -2.76 \nl
N2832 & 96.8$\pm$9.4 & 14.9   & 10.56   & 349$\pm$45  & 123   & -1.94 & 18.00   
& -2.77 \nl
N3115 & 9.8$\pm$0.6  & 2.31   & 1.59    & 278$\pm$36  & 4.74  & -2.53 &  6.15   
& -2.41 \nl
N3377 & 11.6$\pm$0.6 & 0.891  & 0.216   & 131$\pm$17  & 0.713 & -2.47 &  0.176  
& -3.09 \nl
N3379 & 10.8$\pm$0.7 & 1.69   & 0.822   & 201$\pm$26  & 4.29  & -2.29 &  1.33   
& -2.79 \nl
N3608 & 23.6$\pm$1.5 & 2.51   & 1.27    & 206$\pm$27  & 2.87  & -2.64 &  1.49   
& -2.93 \nl  
N4168 & 31.7$\pm$6.2 & 3.28   & 2.22    & 185$\pm$24  & 10.4  & -2.36 &  0.900  
& -3.39 \nl
N4278 & 15.3$\pm$1.7 & 1.90   & 1.25    & 270$\pm$35  & 13.6  & -1.98 &  5.36   
& -2.37 \nl
N4291 & 26.9$\pm$4.1 & 1.65   & 1.11    & 269$\pm$35  & 17.5  & -1.81 &  5.27   
& -2.33 \nl
N4467 & 16.7$\pm$1.0 & 0.0667 & 0.0355  &  87$\pm$11  &\nodata&\nodata&  0.0256 
& -3.14 \nl  
N4472 & 16.7$\pm$1.0 & 10.87  & 8.94    & 273$\pm$36  & 28.5  & -2.51 &  5.65   
& -3.20 \nl
N4473 & 16.1$\pm$1.1 & 1.85   & 0.934   & 188$\pm$25  & 3.48  & -2.46 &  0.971  
& -2.98 \nl
N4486 & 16.7$\pm$1.0 & 9.12   & 9.04    & 345$\pm$45  & 38.6  & -2.36 & 17.04   
& -2.72 \nl
N4486B& 16.7$\pm$1.0 & 0.109  & 0.0358  & 178$\pm$23  & 10.0  & -0.541&  0.750  
& -1.68 \nl
N4552 & 15.7$\pm$1.2 & 2.37   & 1.56    & 269$\pm$35  & 4.79  & -2.52 &  5.27   
& -2.47 \nl
N4564 & 14.9$\pm$1.2 & 0.767  & 0.416   & 153$\pm$20  & 2.48  & -2.24 &  0.367  
& -3.05 \nl
N4594 & 9.9$\pm$0.9  & 5.10   & 3.08    & 248$\pm$32  & 7.39  & -2.65 &  3.59   
& -2.93 \nl
N4621 & 18.6$\pm$1.9 & 4.07   & 2.34    & 222$\pm$29  & 3.39  & -2.84 &  2.13   
& -3.04 \nl
N4636 & 15.0$\pm$1.1 & 3.83   & 3.16    & 180$\pm$23  & 2.22  & -3.15 &  0.791  
& -3.60 \nl
N4649 & 17.3$\pm$1.3 & 7.85   & 6.00    & 331$\pm$43  & 44.3  & -2.14 & 14.02   
& -2.63 \nl
N4660 & 13.2$\pm$1.3 & 0.226  & 0.118   & 211$\pm$28  & 2.41  & -1.70 &  1.67   
& -1.85 \nl
N4874 & 100.9$\pm$9.8& 26.1   & 22.3    & 230$\pm$30  & 225   & -2.00 &  2.51   
& -3.95 \nl
N4889 & 91.6$\pm$8.9 & 18.2   & 11.9    & 373$\pm$49  & 265   & -1.68 & 24.63   
& -2.68 \nl
N6166 & 131$\pm$13   & 28.2   & 18.6    & 311$\pm$41  & 330   & -1.75 & 10.45   
& -3.25 \nl
N7332 & 23.5$\pm$2.3 & 1.05   & 0.193   & 125$\pm$16  &\nodata&\nodata&  0.141  
& -3.14 \nl
N7768 & 114$\pm$11   & 15.62  & 9.79    & 224$\pm$29  & 101   & -2.03 &  2.22   
& -3.64 \nl   
\enddata  
\tablenotetext{}{NOTE.--Distances in Mpc.
$L_{\rm bulge}$ in $10^{10}L_{\odot}$.
$M_{\rm bulge}$ in $10^{11}M_{\odot}$.
$\sigma_c$ in \kms.
$M_{\rm fit}$ and $M_{\bullet}$ in $10^8 M_{\odot}$.}
\end{deluxetable}

\begin{figure}
\plotone{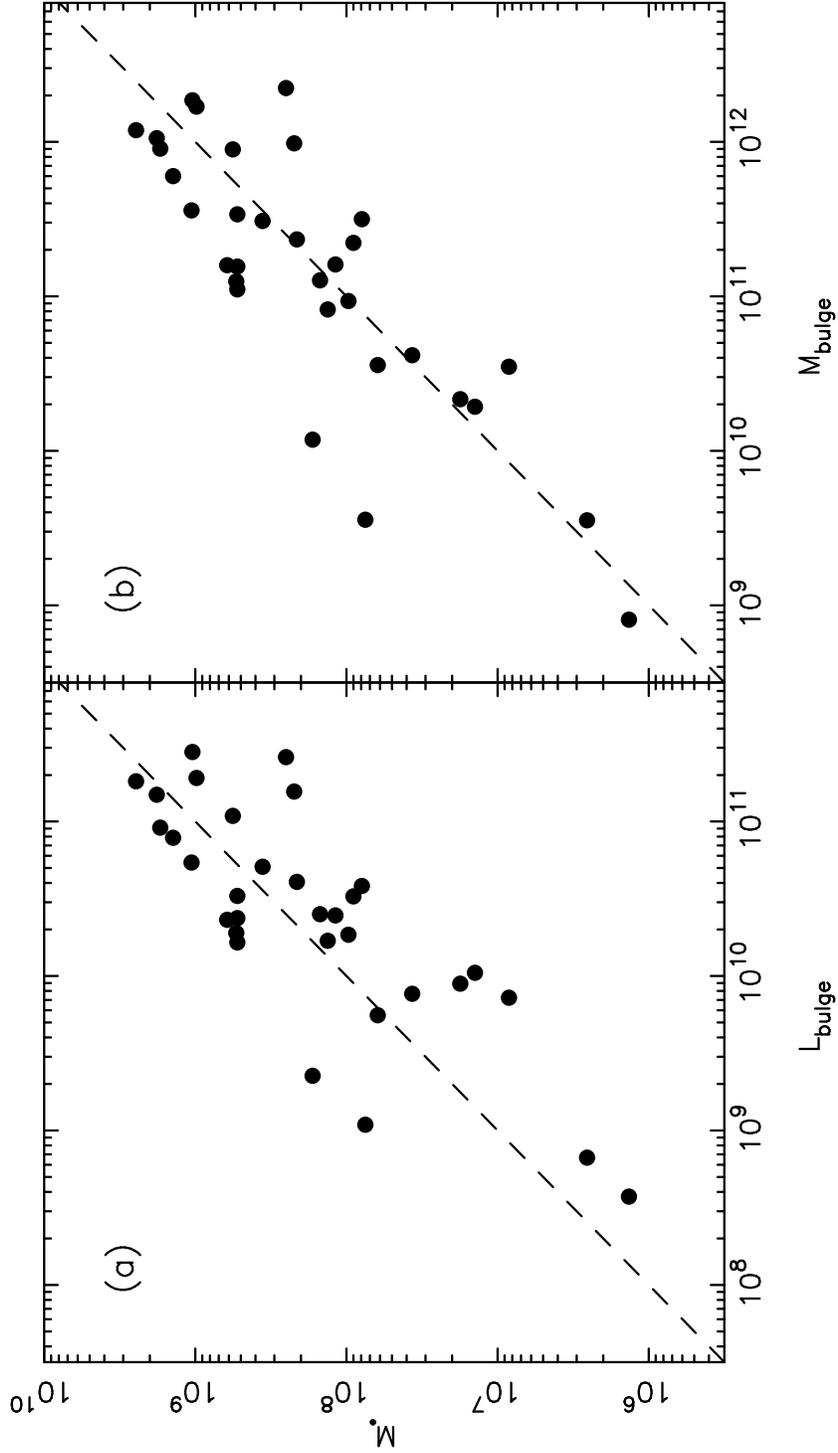} 
\caption{Correlations between black hole mass and:
(a) $V$-band bulge luminosity; (b) bulge mass.
Masses are in units of solar masses and luminosities
in solar luminosities.
Dashed lines are 
$\mh/~{\rm M}_{\odot}=10^{-2} L_{\rm bulge}/L_{\odot}$
(left panel) and 
$\mh/~{\rm M}_{\odot}=10^{-3} M_{\rm bulge}/M_{\odot}$
(right panel).
}
\end{figure}

\begin{figure}
\plotone{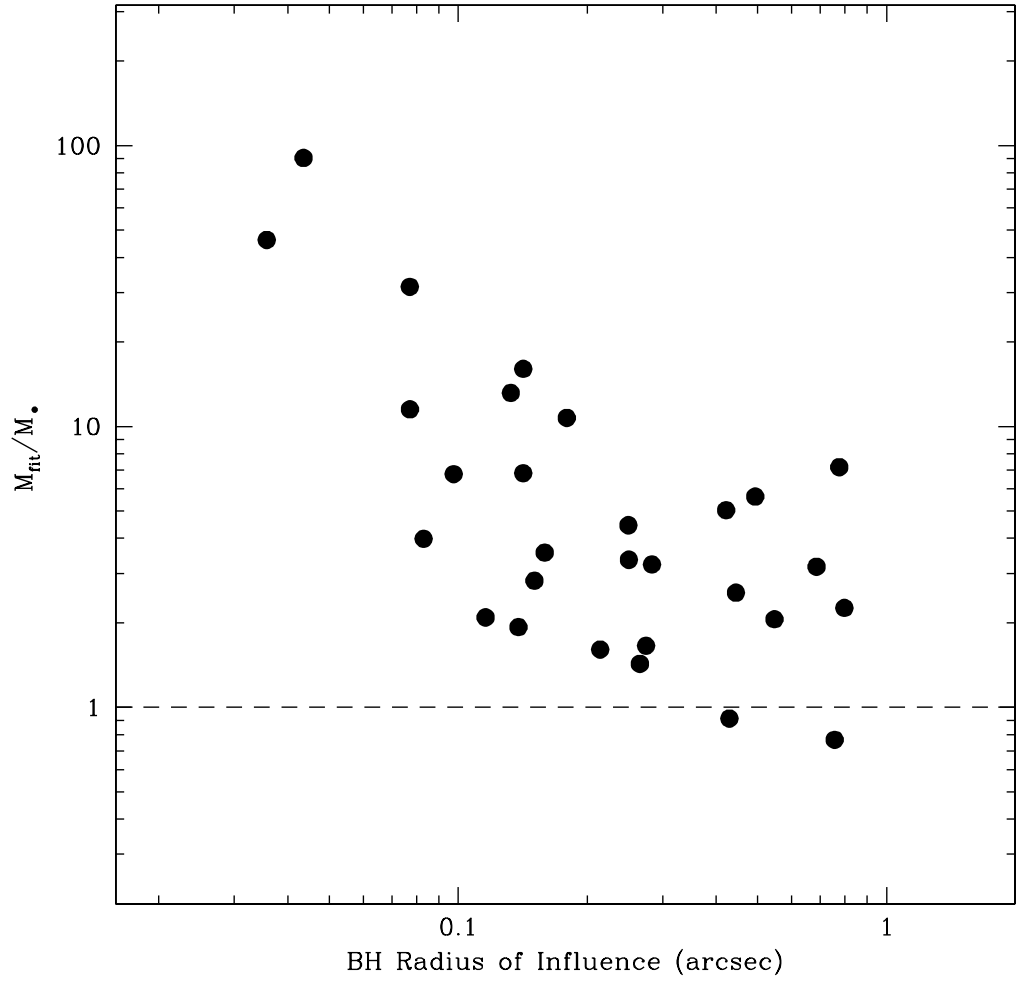} 
\caption{Ratio of black hole mass computed by Magorrian et al.
(1998), $M_{\rm fit}$,
to black hole mass computed from the $\ms$ relation, 
$\mh$, as a function of the radius of influence of the 
nuclear BH.
}
\end{figure}

\begin{figure}
\plotone{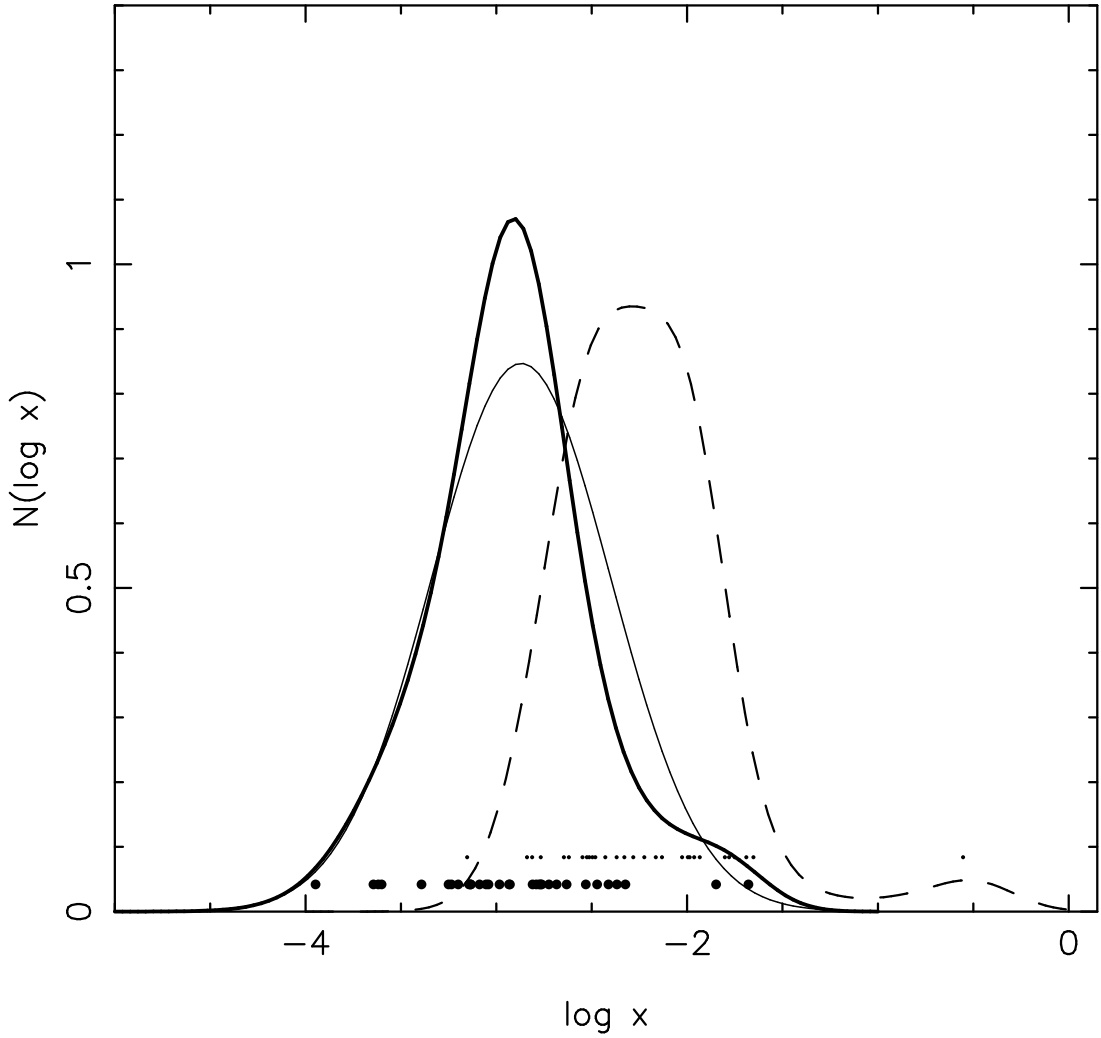}
\caption{Frequency function of $\log x$ where $x=\mh/\mb$.
Heavy solid line was derived from BH masses computed via the 
$\ms$ relation, equation (1); data are shown as the large
dots.
Dashed line was derived using the Magorrian et al. black
hole masses; data are shown as the small dots.
Thin solid line is the best-fit Gaussian approximation 
to $N(\log x)$.
Each curve assumes a measurement uncertainty in $\log x$ of
$0.15$.
}
\end{figure}

\begin{figure}
\plotone{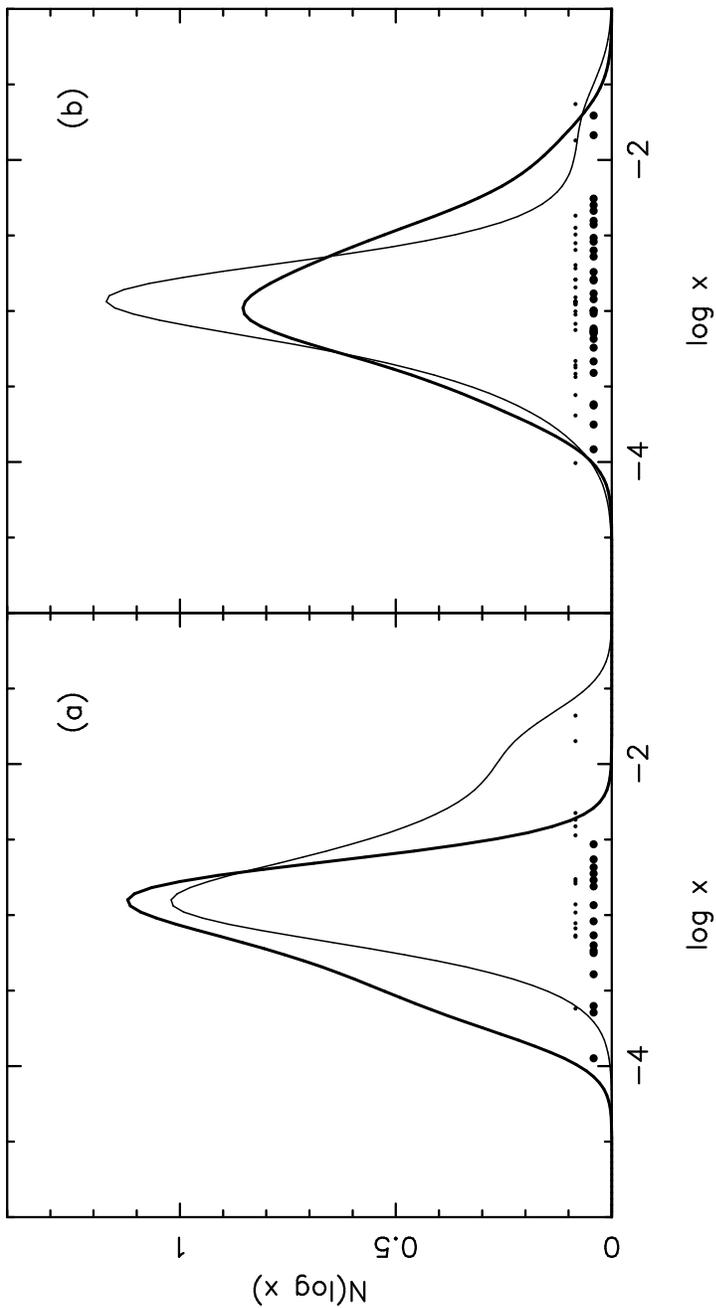}
\caption{
(a) $\hat N(\log x)$ computed separately for the high-$\mb$
(thick line) and low-$\mb$ (thin line) galaxies in Table 1.
(b) Effect of varying the assumed slope of the $\ms$ relation.
Thick line: $\alpha=5.25$.
Thin line: $\alpha=3.75$.
Each curve assumes a measurement uncertainty in $\log x$ of
$0.15$.
}
\end{figure}

\end{document}